\documentclass[aps,twocolumn]{revtex4}
\begin{document}

\title{Position Space Versions of Magueijo-Smolin Doubly Special Relativity 
Proposal and the Problem of Total Momentum.}

\author{A.A. Deriglazov}
\thanks{\noindent e-mail: alexei@ice.ufjf.br}
\author{B. F. Rizzuti}
\address{Dept. de Matematica, ICE, Universidade Federal de Juiz de Fora,\\
Juiz de Fora, MG, Brasil.}

\begin{abstract}
\noindent
We present and discuss two different possibilities to construct position space 
version for Magueijo-Smolin (MS) doubly special relativity proposal. The first 
possibility is to start from ordinary special relativity and then to define 
conserved momentum in special way. It generates MS invariant 
as well as nonlinear MS transformations on the momentum space, 
leading to consistent picture for one-particle sector of the theory. 
The second possibility is based on the following observation. Besides the  
nonlinear MS transformations, the MS energy-momentum relation is invariant also 
under some inhomogeneous linear transformations. 
The latter are induced starting from linearly realized Lorentz group in 
five-dimensional position space. Particle dynamics and kinematics are 
formulated starting from the corresponding five-dimensional interval. 
There is no problem of total momentum in the theory. 
The formulation admits two observer independent scales, the speed of light, $c$, 
and $k$ with dimension of velocity. We speculate on different possibilities 
to relate $k$ with fundamental constants. In particular, expression of 
$k$ in terms of vacuum energy suggests emergence of (minimum) quantum of mass. 
\end{abstract}

\maketitle

{\bf PAC codes:} 98.80 Cq 

\section{Introduction}

It has been discussed in a number of papers (see [1-3] and references therein)
that some experimental data at higher energies may be considered as 
pointing on deviations from special relativity kinematics. To describe the deviations, 
one can add rotationally invariant (but not Lorentz invariant) interaction terms 
with some tiny parameters to the standard model Lagrangian. It leads, 
in particular, to deformation of the special relativity energy-momentum relation [1]. 
There are at least two possibilities to treat the situation. 

A. The noninvariant terms indicates on existence of preferred frame (the frame in 
which the cosmic microwave background is isotropic, so the interaction is 
rotationally invariant [1]), thus the Lorentz symmetry is violated at higher 
energies. 

B. Other point of view, which implies preservation of the principle of relativity 
of inertial frames, has been put forward in the works [3-5].
The underlying symmetry group remains, as before, the Lorentz group, but it's  
realization on the momentum space is supposed to be different from the ordinary one. The deformed 
realization involves tiny parameter in such a way, that one recovers the standard 
Lorentz transformations in some limit. The parameter turns out to be one more 
observer independent scale of the theory in addition to the speed of light, from 
which it follows the name "doubly (deformed) special relativity" (DSR). 
In particular, the scale introduced in [4, 5] was identified with the Planck 
energy.  

There is a number of attractive motivations for such a kind modification, some 
of them are enumerated below.

1. Mathematical motivation. To relate the ordinary Lorentz group realization 
\begin{eqnarray}\label{0}
x^{'\mu}=\Lambda^{\mu}{}_{\nu}x^{\nu}, 
\end{eqnarray}
with transformations among inertial observers (namely with the boosts), 
one introduces the parameter $c$: $x^0=ct$, which has rather special properties 
in the resulting theory: a) the transformations degenerate in the limit 
$V\rightarrow c$; b) the limit $c\rightarrow\infty$ reproduce the Galileo relativity; 
c) the velocity $v=c$ turns out to be observer independent scale of the theory 
(namely, maximum signal velocity). Then one asks on 
generalization of the special relativity which admits more than one dimensional 
parameter with similar properties. 

2. Physical motivation. Several quantum gravity calculations indicate on 
modification of the special relativity energy-momentum relation (suppressed by the 
Planck scale). Then one can believe on DSR as an intermediate theory where the 
quantum gravity effects are presented even in the regime of negligible gravitational 
field [5, 6]. 

3. "Experimental" motivation. The modified energy-momentum relation implies 
corrections  to the GZK cut-off (or even absence of the latter [1]), so DSR 
proposals may be relevant for discussion of the threshold anomalies in ultra 
high-energy cosmic rays [3, 5]. Possible energy dependence of speed of 
light in DSR theory may be relevant to planned gamma-ray observations 
[5]. Some other astrophysics data were discussed in the DSR framework,  
see [7, 3, 2]. 

The DSR proposals [3-5] have been formulated in space of conserved 
energy-momentum $p^\mu$, that is as a list of kinematical rules of the new theory. 
In particular, Magueijo-Smolin (MS)  
suggestion [4, 5] is to take the momentum space realization of the Lorentz 
group in the form
\begin{eqnarray}\label{00}  
\Lambda_U=U^{-1}\Lambda U, \qquad \qquad 
 p^{'\mu}=(\Lambda_U)^\mu{}_\nu p^{\nu},
\end{eqnarray} 
where $\Lambda$ represents 
ordinary Lorentz transformation and $U(p^\mu, \lambda)$ is some operator which depends 
on the invariant scale $\lambda$.
Ordinary energy-momentum relation $(p^\mu)^2=-m^2$ is not invariant under 
the realization and is replaced by $[U(p^\mu)]^2=-m^2$. It suggests kinematical 
predictions different from that of special relativity. 

Unfortunately, the list of kinematical rules of MS is not complete.  
Central problem of the DSR kinematics is consistent definition of total momentum 
for many particle system (see [5, 8]). Actually, due 
to non-linear form of the 
transformations, ordinary sum of momenta does not transform as the 
constituents. Different covariant composition rules proposed in the literature 
lead to hardly acceptable features [8], in particular, one is faced with the 
"soccer ball problem". One possibility to avoid the problem was proposed in the 
recent work [12]. 

To discuss physical interpretation of the DSR kinematics it is desirable to find 
the underlying space-time version of the theory, that is to construct realization 
of the Lorentz group in position space, and then to formulate dynamical 
problems which generate the DSR kinematics in one or another way. The issue 
turns out to be rather delicate question, as it was discussed in [9-12]. 
In this work we discuss two different possibilities to  construct the 
space-time version of the MS DSR kinematics. To find it one needs, in fact, 
to decide what is the 
relation between $x^\mu$ and the conserved energy-momentum $p^\mu$.  
The relation is absent in the MS construction and can be different now from 
the standard one. In Section 2 we suppose that the underlying version is  
ordinary special relativity and then deform definition of the momentum in an 
appropriate way. It generates the MS invariant 
as well as the nonlinear MS transformations on the momentum space, 
leading to consistent picture for one-particle sector of the theory.
In Section 3 we consider the opposite possibility: one takes the standard definition 
of the momentum and then deforms the Lorentz group realization.
We present linear realization of the Lorentz group in 
five-dimensional position space which leads to the MS energy-momentum relation. 
Particle dynamics and kinematics are 
formulated starting from the corresponding five-dimensional interval. 
We point that there is no problem of total momentum in the theory. 
The formulation admits two observer independent scales, the speed of light, $c$, 
and $k$ with dimension of velocity. In Section 4 we discuss different possibilities 
to identify $k$ with fundamental constants. 

\section{MS Doubly Special Relativity Starting from the Special Relativity}

Imposing that the special relativity can be deformed by the scale 
$\lambda$, one can write for the conserved momentum the expression
$p^\mu=m f^\mu \, _\nu (x^\mu, \dot x^\mu,\lambda)\dot x^\nu$ 
with $f^\mu \, _\nu \stackrel{\lambda \rightarrow 0}{\longrightarrow} 
\delta^\mu \, _\nu $. Assuming this, one can try to reproduce the given kinematics 
starting, for instance, from ordinary special relativity in the position space. 
The resulting theory is discussed in this Section. It will be demonstrated that 
MS one particle kinematics can actually be considered as 
corresponding to the ordinary 
special relativity particle dynamics, if one deforms definition 
of the special relativity conserved momentum in an appropriate way. The MS invariant 
and the corresponding nonlinear transformations are induced from the standard 
ones for $x^\mu$ variables. 

\subsection{Initial MS proposal}

The initial MS proposal [4] implies that all the inertial observers should 
agree to take the 
deformed dispersion relation for the conserved momentum of a particle
\begin{eqnarray}\label{1}
p^2=-m^2c^2(1+\lambda p^0)^2,
\end{eqnarray}
where $\lambda$ is some observer independent scale. 
Then one needs to find realization of the Lorentz group on the momentum space 
which leaves 
Eq.(\ref{1}) covariant. They have suggested the following nonlinear transformations:
\begin{eqnarray}\label{2}
p^{'\mu}=\frac{\Lambda^\mu \, _\nu p^\nu}{1+\lambda(p^0 - \Lambda^0 \, _\nu p^\nu)}.
\end{eqnarray} 
Now, one notes that $p^\mu=m(1+\lambda p^0)\dot x^\mu$, being substituted in 
Eq.(\ref{1}), gives the 
standard relation $(\dot x^\mu)^2=-c^2$. Solution of the equation is
\begin{eqnarray}\label{3}
p^\mu=\frac{m \dot x^\mu}{1-\lambda m \dot x^0}.
\end{eqnarray}
The standard Lorentz transformation (\ref{0}) for $x$ 
generates the transformation (\ref{2})
for $p$ defined in accordance with Eq.(\ref{3}). 

Equations of motion for $x^\mu$ can be restored from Eq.(\ref{1}) and the 
condition of momentum conservation $\dot p^\mu=0$ 
\begin{eqnarray}\label{4}
S^\mu\equiv(\frac{\dot x^\mu}{1-\lambda m \dot x^0})^{.}=0, \qquad
\dot x^\mu \dot x_\mu=-c^2.
\end{eqnarray}
It gives space-time version of (\ref{1}), (\ref{2}) with the conserved momentum 
defined by Eq.(\ref{3}).
Let us point that the first equation in (\ref{4}) is covariant with respect 
to (\ref{0}): $S^\mu(x^{'})=B^\mu{}_\nu S^\nu(x)$, with $B$-matrix being invertible 
on the same region as 
(\ref{3}). The system (\ref{4}) describes the free particle motion, which can be 
confirmed by direct computation (in fact, the first equation in 
(\ref{4}) can be substituted by $\ddot x^\mu=0$). 

\subsection{General MS proposal}

In general case (\ref{00}) one takes some particular form of the operator 
$U(p, \lambda)$ to obtain the deformed energy-momentum relation and the 
corresponding Lorentz group realization in the momentum space.
It is not difficult to generate these models by using of the trick described 
in the previous subsection, that is starting from the special relativity particle 
dynamics. Let us consider first the polynomial energy-momentum relation 
\begin{eqnarray}\label{5}
p^\mu p_\mu=-m^2 c^2(1+ \sum^N _{n=1} \alpha_n (\lambda p^0)^n)^2.
\end{eqnarray}
One writes the equations for determining the momenta in terms of $\dot x^\mu$
\begin{eqnarray}\label{6}
p^0=m (1+ \sum^N _{n=1} \alpha_n (\lambda p^0)^n)\dot x^0, \cr
p^i= m (1+ \sum^N _{n=1} \alpha_n (\lambda p^0)^n)\dot x^i.
\end{eqnarray}
Assuming that the first equation can be resolved in relation of $p^0$, one finds 
solution of the system (\ref{6}) in the form 
\begin{eqnarray}\label{7}
p^0 = p^0(\dot x^0, \lambda), \qquad \qquad 
p^i = p^i(\dot x^i, \dot x^0, \lambda).
\end{eqnarray}
Then the standard Lorentz transformations for $x^\mu$ (\ref{0}) generate some nonlinear  
realization of the Lorentz group on the momentum space (\ref{7}). By construction, 
the dispersion relation (\ref{5}) is invariant with respect to this realization. 

In general case the energy-momentum relation is of the form [5]
\begin{eqnarray}\label{8}
(p^0)^2f^2_1(p^0, \lambda)-(p^i)^2f^2_2(p^0, \lambda)=m^2c^2. 
\end{eqnarray}
Equations for determining the momenta acquire the form
$p^0f_1(p^0, \lambda)$ $=m\dot x^0, ~  p^if_2(p^0, \lambda)=m\dot x^i$. Assuming that 
the first equation can be resolved in relation of $p^0$, one finds 
solution of the system in the form (\ref{7}).  

Thus we have demonstrated that MS construction can be considered as corresponding to 
space-time particle dynamics of the special relativity, that is the kinematics is 
generated starting from $\dot x^2=-c^2$, ~  $\ddot x^\mu=0$, ~ 
$x^{'\mu}=\Lambda^\mu\,_\nu x^\nu$, which gives  a consistent picture in one-particle 
sector.     
In ordinary special relativity, expression for the  
conserved momentum is dictated by the translation invariance. Then, how one can take 
some different expression of the type (\ref{3})? The point is that the MS kinematics 
is well 
defined for the one particle sector only. In this case definition of conserved 
momentum is conventional (since $\dot x^\mu=const$, any function of $\dot x^\mu$ 
can be taken as the conserved quantity).
For many-particle case, definition of the conserved momentum is not conventional. 
On absence of consistent addition rule in many-particle sector of the MS kinematics, 
the presented point of view, being quite simple, seems to be unreasonable. 
On this reasoning we propose below other position version, which generates 
the MS invariant on momentum space. In particular,
the version turns out to be free of the problem of total momentum. 

\section{Doubly Special Relativity as $M^4\times\Re$ "Galilean" Relativity}

Discussion of this Section is based on the following two observations.

1. According to the previous analysis, the problem of total momentum is due to the 
fact that the conserved momentum, being defined in MS DSR  
in terms of the velocities in the nonlinear form (\ref{3}), is not a tangent vector. 
To avoid the problem, 
one preserves the ordinary definition of the momentum, deforming the space-time 
interval in such a way that the resulting theory reproduce the MS invariant 
(\ref{1}) in the momentum space. Analysis of the corresponding particle dynamics 
reveals that the 
evolution parameter $T$ of the model does not coincide with proper 
time of the particle, see Subsection 3.1. 
So, in this framework, the deformed dispersion relation 
suggests emergence of one more evolution parameter in the position version of DSR 
theory. One has $x^0, ~ \tau, ~ T$, instead of $x^0, ~ \tau$ of the special 
relativity theory. 
It prompts that the appropriate group-theoretic framework for the model may be 
some realization of the Lorentz group in 5-dimensional position space. 

2. Besides the nonlinear symmetry (\ref{2}), the MS relation (\ref{1}) is invariant 
also under the linear inhomogeneous transformations
\begin{eqnarray}\label{9}
p^{'0}=\Lambda^0 \, _0 p^0 + \frac{\Lambda^0 \, _i p^i}{\sqrt{1-c^2 \lambda^2}}-
(1- \Lambda^0 \, _0)\frac{c^2 \lambda}{1-c^2 \lambda^2}\cr
p^{'i}=\sqrt{1-c^2 \lambda^2}\Lambda^i \, _0 p^0 + \Lambda^i \, _j p^j + 
\Lambda^i \, _0 \frac{c^2 \lambda}{1-c^2 \lambda^2}. 
\end{eqnarray}
It suggests linear realization of the Lorentz group in the position space. We consider below the  
position version which leads to momentum transformations with more symmetric structure, 
see Eqs.(\ref{13}), (\ref{22}). In contrast to the MS case, the resulting model 
has no degeneration in the limit of a particle with large mass. Generalization on 
the MS case is straightforward and will be presented in a separate work.

\subsection{Particle dynamics based on the MS invariant}

Assuming the standard relation among  momenta and velocities 
$p^\mu=m\frac{dx^\mu}{dT}$, the equations of motion
\begin{eqnarray}\label{11}
\frac{d^2x^\mu}{dT^2} \equiv \ddot x^\mu=0, \qquad
\dot x^\mu \dot x_\mu=-c^2(1-\lambda mc  \dot x^0)^2,
\end{eqnarray}
corresponds to the MS kinematics. Here T is invariant evolution parameter. Analysis 
of the dynamics is similar to that of ordinary relativistic particle [13]. One 
takes the deformed relation $dx^0=c(1-\lambda^2m^2c^4)^{-\frac{1}{2}}dt$,  
otherwise the invariant velocity scale depends on the rest mass. 
The relation degenerates for $m^2\rightarrow \frac{1}{\lambda^2c^4}$, which represents 
the position version of the "soccer ball problem" (the model considered below is free 
of this problem).
Solution of Eq.(\ref{11}) is 
$x^i(t)=v^it+a^i$, where the three-velocity is restricted by $\mid v^i\mid\le c$. 
From the second equation in (\ref{11}) it follows, that the maximum velocity $c$ 
turns out to be the invariant scale: if a particle 
has $\mid v^i\mid=c$ for one observer, it has the same velocity for any 
other observer. The same equation allows one to relate proper time of the 
particle $dt$ with the evolution parameter 
\begin{eqnarray}\label{12}
dT=\sqrt{\frac{1+\lambda mc^2}{1-\lambda mc^2}}dt.
\end{eqnarray}
As it was mentioned above, the proper time does not coincide with the evolution 
parameter, similarly to geodesic particle motion in gravity theory. 

\subsection{Realization of the Lorentz group on $M^4\times\Re$-space}
Let us consider a realization of the Lorentz group 
$\left\{\Lambda^\mu \,_\nu\right\}$ 
in $5$-dimensional position space $M^4\times\Re$ (parameterized by $(x^\mu, ~ x^5)$, 
$\eta_{\mu\nu}=(- + + +)$): $x^{'A}=T^{A}\,_{B}(\Lambda)x^{B}$, namely  
\begin{eqnarray}\label{13}
x^{'\mu}= \Lambda^\mu \, _\nu x^\nu +(\delta^\mu \, _0 - \Lambda^\mu \, _0)x^5,\cr 
x^{'5}=x^{5}. \qquad \qquad \qquad 
\end{eqnarray}
It leaves $x^5$ invariant. Let us enumerate some properties of the DSR theory based 
on Eq.(\ref{13}). 
\par
1- Let us note that deformations of the special relativity 
in some domain by means of the transformation $\Lambda_{def}=U^{-1}\Lambda U$ suggest 
existence of (singular) change of variables $x_{SR}=U^{-1}x$. The variable $x_{SR}$ 
has the standard 
transformation law under $\Lambda_{def}$: $x^{'}_{SR}=\Lambda x_{SR}$. 
It is true for the 
Fock-Lorentz realization [14] and for the recent DSR proposals [9, 12]. Moreover, 
different DSR proposals in momentum space can be considered either as different 
definitions of the 
conserved momentum $p^\mu$ in terms of the de Sitter momentum space 
variables $\eta^A$ [8, 11], or as different definitions of $p^\mu$ in terms of the 
special 
relativity velocities $v^\mu=\frac{dx^\mu}{d\tau}$, see Section 2. Thus the known DSR 
proposals state, in fact, that experimentally measurable coordinates can be different 
from the ones specified as "measurable" by the special relativity theory.  
For the case under consideration, Eq.(\ref{13}) suggests, in fact,  
redefinition of the special relativity observer time $x^0_{SR}=x^0-x^5$.
\par
2- Eq.(\ref{13}) is realization of the Lorentz group: $T(\Lambda_1)T(\Lambda_2)=
T(\Lambda_1\Lambda_2)$, with $det T\ne 0$.
\par 
3- The transformations with $\Lambda^\mu \, _\nu=( \Lambda^0 \, _0 =1, ~  
\Lambda^0 \, _i= \Lambda^i \, _0 =0, ~ \Lambda^i \, _j\equiv R^i \, _j, ~ 
R^T=R^{-1})$ 
are identified with space rotations of $M^4$: 
$x^{'i}=R^i \, _j x^j, ~ ~   x^{'0}=x^0, ~ ~ T^{'}=T$. They are not deformed by 
means of $x^5$.  
\par
4- One notes similarity of Eq.(\ref{13}) with the Galileo transformations in 
$\Re^3 \times \Re$:
$x^{'i}=G^i \, _j x^j - V^i t, ~ ~ ~ t=t^{'}$, 
but in (\ref{13}) it was taken the Minkowski space $M^4$ instead of $\Re^3$.
\par
5- To make the physical interpretation of Eq.(\ref{13}), one assumes that it 
describes the transformation law among inertial observers. To construct the 
covariant dynamics, all the observers should agree on taking $x^5$ as the 
evolution parameter, the latter is independent on a frame in accordance 
with Eq.(\ref{13}). Thus one takes $x^5=kT$, where $[T]=sec$ and $k$ is some 
constant $[k]=\frac{m}{sec}$ (the notations $\dot x=\frac{dx}{dT}$ will be 
used below). The special relativity then corresponds to 
the limit $k\to 0$ in Eq.(\ref{13}). Note that the choice $x^5=cT$ is not 
interesting in this respect. The additional dimension $x^5$ decouples 
from $M^4$ in the limit.  
\par
6- The invariant interval of the transformations (\ref{13}) is 
\begin{eqnarray}\label{14}
-ds_5^2=\eta_{\mu\nu}dx^\mu dx^\nu+(dx^5-2dx^0)dx^5,
\end{eqnarray}
and leads to deformed energy-momentum relation in the momentum 
space, see Eq.(\ref{21}) below. The MS relation corresponds 
to the interval $\eta_{\mu\nu}dx^\mu dx^\nu+(dx^5+dx^0)^2$.
\par
7- Similarly to the Magueijo-Smolin [4], let us ask on existence of the tangent 
space vector with zero component preserved by the modified action of the Lorentz 
group. From Eq.(\ref{13}) and from the requirement 
$\dot x^{'0}=\dot x^0$ one finds the unique solution $\dot x^\mu=(k,0,0,0)$. 
Thus $k$ represents some observer independent scale of the model. 

To conclude this subsection, we point that the conserved momentum can be defined 
in the standard form
\begin{eqnarray}\label{15}
p^A=m\dot x^A=(m\dot x^{\mu}, ~ mk),  ~ \Longrightarrow ~    
p^{'A}=T^A\,_B(\Lambda)p^B
\end{eqnarray}
Then $P^A=p^A+\pi^A$ is the covariant composition rule 
($P$ obeys the Eq.(\ref{15})),  
and $P^0$ is unbounded from above, in contrast with the MS theory. 
The model turns to be free of the problem of total momentum.

\subsection{Covariant dynamics of $M^4\times\Re$ particle}
To describe the free particle motion, one takes $\frac{d^2x^\mu}{dT^2}=
\ddot x^\mu=0$. Similarly to the special 
relativity theory, one needs to specify the relation among $T$ and the observer time 
$x^0$. It can be achieved by using of the invariant interval (\ref{14}) 
which we have in our disposal. 
Since the interval involves one more variable, the latter must be specified also. 
Let us write Eq.(\ref{14}) in the form 
\begin{eqnarray}\label{16}
\eta_{\mu\nu}\dot x^\mu \dot x^\nu=-\left(\dot s_5^2-k^2+2k\dot x^0\right).
\end{eqnarray}
To obtain the special relativity in the limit $k\to 0$ one takes 
$\dot s_5^2=f(c, k)$, where $f\to c^2$ when $k\to 0$. One notes that the equation 
for $\dot s_5^2$ is 
invariant with respect to the transformations (\ref{13}). In particular, taking 
$\dot s_5^2=c^2+k^2$ 
one obtains the following relativistic dynamics
\begin{eqnarray}\label{17}
\ddot x^\mu=0, \qquad 
\eta_{\mu\nu}\dot x^\mu \dot x^\nu=-c^2\left(1+
\frac{2k}{c^2}\dot x^0\right). 
\end{eqnarray}
Analysis of the dynamics is similar to that of ordinary relativistic particle 
[13]. One notes that $\ddot x^0=0$ is consequence of other equations in 
(\ref{17}) for $\dot x^0\ne k$. The Cauchy problem for the remaining equations is 
$x^i(T=0)=a^i, ~ x^0(T=0)=0, ~ \dot x^i(T=0)=b^i$, for any $a^i, ~ b^i$, where 
the clocks synchronization is implied. One finds the solution
\begin{eqnarray}\label{18}
x^i=b^iT+a^i, \qquad x^0=(k+\sqrt{k^2+c^2+(b^i)^2})T.
\end{eqnarray}
Assuming that these expressions describe a particle motion in parametric form, 
one finds, taking the standard relation $dx^0=cdt$  
\begin{eqnarray}\label{19}
x^i(t)=v^it+a^i, \qquad v^i=\frac{cb^i}{k+\sqrt{k^2+c^2+(b^i)^2}}.
\end{eqnarray}
where $v^i$ represents the initial three-velocity. The latter is restricted: 
$\mid v^i\mid\rightarrow c$ when $\mid b^i\mid\rightarrow\infty$.
From the second equation in (\ref{17}) one finds, that the maximum velocity $c$ 
turns out to be the invariant scale: if a particle 
has $\mid v^i\mid=c$ for one observer, it has the same velocity for any 
other observer. The same equation allows one to relate proper time of the 
particle with the evolution parameter 
\begin{eqnarray}\label{20}
dT=\left[\sqrt{1+\left(\frac{k}{c}\right)^2}-\frac{k}{c}\right]dt,
\end{eqnarray}
where the proportionality factor is non degenerated.

In the momentum space one finds the deformed energy-momentum relation 
\begin{eqnarray}\label{21}
p^2=-m^2c^2\left(1+
\frac{2k}{mc^2}p^0\right), 
\end{eqnarray}
which is invariant under the transformations generated by (\ref{14})
\begin{eqnarray}\label{22}
p^{'\mu}=\Lambda^\mu{}_\nu p^\nu+(\delta^\mu{}_0-\Lambda^\mu{}_0)mk.
\end{eqnarray}

\section{Discussion: Identifying $k$ with Fundamental Constants}

In this work we have proposed and discussed two different position space 
versions for the MS DSR kinematics. One possibility is to start from ordinary 
special 
relativity and then to define the conserved energy and momentum in special  
way (see Eq.(\ref{3}) for the initial MS proposal). It generates the MS invariant 
(\ref{1}) as well as the MS transformations (\ref{2}) on the momentum space, 
leading to consistent 
picture for one-particle sector of the theory. Generalization on multi-particle 
sector is problematic, mainly due to the fact that it is not yet known a 
consistent rule for addition of momenta [8]. 

The problem of total momentum can be avoided, if one preserves ordinary definition 
for the conserved momentum. It implies deformation of the special relativity 
interval as well as the Lorentz group realization in the position space. 
Following this line, we have presented linear realization of the Lorentz 
group in five-dimensional position space (\ref{13}), with the fifth coordinate being 
invariant under the transformations. Particle dynamics and kinematics were 
formulated starting from the five-dimensional interval. In particular, the model 
leads to the MS-type energy-momentum relation (\ref{21}). 
This proposal can be compared with interpretation of DSR kinematics in terms of 
pentamomentum of (Anti)de Sitter space [16]. Due to linear realization of the Lorentz group  
on the space, the sum rule for composite system is simply add the pentamomentum of the 
constituents. It leads to rescaling of $k$ for composite systems and reproduces the 
sum rule for quadrimomentum proposed in [5]. This interpretation implies also a concept of 
five-dimensional spacetime, with the fifth coordinate being characteristic of a 
reference frame mass [16].     

The position version constructed in Section 3 naturally leads to two observer 
(and position) independent scales with dimension of velocity: $c$ and $k$. 
The scale $k$ is supposed to be small, and to obtain the special relativity 
in the limit $k\rightarrow 0$ one identifies $c$ with speed of light. 
Now one asks on  
interpretation of $k$ in terms of other fundamental constants. There are three 
dimensional constants which  
should play a fundamental role in the quantum theory of gravity: speed of light $c$, 
Newton's gravitational constant $G$, and the Planck constant $\hbar$. They can be 
used to construct the Planck scales, of length 
$l_p=\sqrt{\frac{\hbar G}{c^3}}\sim 10^{-35} m$ 
(or energy $E_p=\sqrt{\frac{\hbar c^5}{G}}\sim 10^{8} \frac{kg m^2}{seg^2})$, 
time $t_p=\sqrt{\frac{\hbar G}{c^5}}\sim 10^{-43} seg$, 
and mass $M_p=\sqrt{\frac{\hbar c}{G}}\sim 10^{-8} kg$. 
Besides this, a possibility to include some other dimensional scales have been discussed 
[7, 15, 12]. In particular, in the work [7] it was analysed an algebraic construction which 
implies three scales $c, ~ E_p, ~ \Lambda$, with $E_p$ identified with the Planck energy 
and $\Lambda$ being the cosmological constant. The latter appears also in a natural way 
in position version of the DSR theory based on conformal group [12]. So, we include the 
cosmological constant in our subsequent analysis. 
  
Let us discuss different possibilities to relate $k$ with the fundamental 
constants. 
\par
1. On dimension grounds, one writes the scale $k$ in terms of the Planck energy 
$k=\frac{mc^3}{2E_p}$. It implies the energy-momentum relation 
$p^2=-m^2c^2(1+\frac{cp^0}{E_p})$,
as well as dependence of the Lorentz boosts (\ref{13}) on rest mass
\begin{eqnarray}\label{220}
x^{'\mu}= \Lambda^\mu \, _\nu x^\nu +(\delta^\mu \, _0 - \Lambda^\mu \, _0)
\frac{mc^2}{2E_p}cT.
\end{eqnarray}
The value of $k$ for the proton is of order $10^{-11}$. For a body with rest  
mass being of order of the Planck mass, the DSR results diverge from that of 
special relativity. Thus, one is faced with the "soccer ball problem".
\par
2. By using of the cosmological constant it is possible to avoid appearance of 
the mass in the transformation low, one takes $k=\sqrt[3]{\Lambda G \hbar}$.
For the case, $k$ is of order $10^{-32}$. Deviation from the special relativity 
dynamics does not depend on the rest mass, see Eq.(\ref{17}).
\par
3. One more possibility is $k=\frac{E_{vac}}{mc}$, where the vacuum energy is $E_{vac}=\Lambda\sqrt{\frac{\hbar^3G}{c}}$. One has 
\begin{eqnarray}\label{23}
p^2=-m^2c^2(1+2\frac{E_{vac}p^0}{m^2c^3}),
\end{eqnarray}
\begin{eqnarray}\label{24}
x^{'\mu}= \Lambda^\mu \, _\nu x^\nu +(\delta^\mu \, _0 - \Lambda^\mu \, _0)
\frac{E_{vac}}{mc^2}cT.
\end{eqnarray}
This version of the theory approaches to the special relativity for macroscopic 
bodies. The transformations (\ref{24}) suggest a lower bound for an  
energy of a point under observation, that is minimum quantum of mass.   

\noindent ACKNOWLEDGMENTS:
One of the authors (AAD) would like to thank the Brazilian Research Agencies  
CNPq and FAPEMIG for financial support.

\end{document}